\documentclass[reprint,amsmath,amssymb,aps,pra,longbibliography]{revtex4-1}

\usepackage[margin=0.75in, top=1in, bottom=0.85in]{geometry}

\usepackage[final,letterspace=-0]{microtype}
\usepackage[colorlinks=true, linkcolor=blue, urlcolor=blue, citecolor=blue, anchorcolor=blue]{hyperref}

\usepackage{mathpazo} 
\usepackage{upgreek}
\usepackage{xcolor}

\usepackage[]{graphicx}
\usepackage{dcolumn}
\usepackage{bm}
\usepackage{helvet}

\begin{document}

\title{Optical space-time wave packets having arbitrary group velocities in free space}

\author{H. Esat Kondakci and Ayman F. Abouraddy}
\affiliation{CREOL, The College of Optics \& Photonics, University of Central Florida, Orlando, Florida 32816, USA}


\begin{abstract} 
{Controlling the group velocity of an optical pulse typically requires traversing a material or structure whose dispersion is judiciously crafted. Alternatively, the group velocity can be modified in free space by spatially structuring the beam profile, but the realizable deviation from the speed of light in vacuum is small. Here we demonstrate precise and versatile control over the group velocity of a propagation-invariant optical wave packet in free space through sculpting its spatio-temporal spectrum. By jointly modulating the spatial and temporal degrees of freedom, arbitrary group velocities are unambiguously observed in free space above or below the speed of light in vacuum, whether in the forward direction propagating away from the source or even traveling backwards towards it.}
\end{abstract}

\maketitle


\noindent
The publication of Einstein's seminal work on special relativity initiated an investigation of the speed of light in materials featuring strong chromatic dispersion \cite{Brillouin60Book}. Indeed, the group velocity $v_{\mathrm{g}}$ of an optical pulse in a resonant dispersive medium can deviate significantly from the speed of light in vacuum $c$, without posing a challenge to relativistic causality when $v_{\mathrm{g}}\!>\!c$ because the information speed never exceeds $c$ \cite{Brillouin60Book,Schulz69ProcIEEE}. Modifying the \textit{temporal} spectrum in this manner is the basic premise for the development of so-called `slow light' and `fast light' \cite{Boyd09Science} in a variety of material systems including ultracold atoms \cite{Hau99Nature}, hot atomic vapors \cite{Kash99PRL,Wang00Nature}, stimulated Brillouin scattering in optical fibers \cite{Song05OE}, and active gain resonances \cite{Casperson71PRL,Gehring05Science}. Additionally, nanofabrication yields photonic systems that deliver similar control over the group velocity through structural dispersion in photonic crystals \cite{Baba08NP}, metamaterials \cite{Dolling05Science}, tunneling junctions \cite{Steinberg93PRL}, and nanophotonic structures \cite{Tsakmakidis17Science}. In general, resonant systems have limited spectral bandwidths that can be exploited before pulse distortion obscures the targeted effect, with the pulse typically undergoing absorption, amplification, or temporal reshaping, but without necessarily affecting the field spatial profile.

In addition to temporal spectral modulation, it has been recently appreciated that structuring the \textit{spatial} profile of a pulsed beam can impact its group velocity in free space \cite{Giovannini15Science,Bouchard16Optica,Lyons18Optica}. In a manner similar to pulse propagation in a waveguide, the spatial spectrum of a structured pulsed beam comprises plane-wave contributions tilted with respect to the propagation axis, which undergo larger delays between two planes than purely axially propagating modes. A large-area structured beam (narrow spatial spectrum) can travel for longer distances before beam deformation driven by diffraction and space-time coupling, but its group velocity deviates only slightly from $c$; whereas a narrow beam deviates further from $c$, but travels a shorter distance. Consequently, $v_{\mathrm{g}}$ is dependent on the size of the field spatial profile, and the maximum group delay observable is limited by the numerical aperture. Only velocities slightly lower than $c$ ($\approx\!0.99999c$) have been accessible in the experiments performed to date with maximum observed group delays of $\sim30$~fs, corresponding to a shift of $\sim10$~$\mu$m over a distance of 1~m (or 1 part in $10^{5}$). 

Here we demonstrate precise and versatile control over the magnitude \textit{and} sign of the free-space group velocity of a propagation-invariant wave packet by sculpting its \textit{spatio-temporal} profile. Instead of manipulating separately the field spatial \textit{or} temporal degrees of freedom and attempting to minimize unavoidable space-time coupling, we intentionally introduce into the wave packet tight spatio-temporal spectral correlations that result in the realization of  arbitrary group velocities: superluminal, luminal, or subluminal, whether in the forward direction propagating away from the source or in the backward direction traveling toward it. The group velocity here is the speed of the wave packet central spatio-temporal peak. By judiciously associating each wavelength in the pulse spectrum with a particular transverse spatial frequency, we trace out a conic section on the surface of the light-cone while maintaining a linear relationship between the axial component of the wave vector and frequency. The slope of this linear relationship dictates the wave packet group velocity, and its linearity eliminates any additional dispersion terms. The resulting wave packets propagate free of diffraction \textit{and} dispersion \cite{Longhi04OE,Saari04PRE,Turunen10PO,FigueroaBook14}, which makes them ideal candidates for unambiguously observing group velocities in free space that deviate substantially from $c$.

There have been previous efforts directed at the synthesis of optical wave packets endowed with spatio-temporal correlations. Several strategies have been implemented to date, which include exploiting the techniques associated with the generation of Bessel beams, such as the use of annular apertures in the focal plane of a spherical lens \cite{Saari97PRL} or utilizing axicons \cite{Alexeev02PRL,Bonaretti09OE,Bowlan09OL} synthesis of X-waves \cite{Lu92IEEEa} during nonlinear processes such as second-harmonic generation \cite{DiTrapani03PRL} or laser filamentation \cite{Faccio06PRL,Faccio07OE} or through direct filtering of the requisite spatio-temporal spectrum \cite{Dallaire09OE,Jedrkiewicz13OE}. The reported \textit{superluminal} speeds achieved with these various approaches in free space have been to date $1.00022c$ \cite{Bonaretti09OE}, $1.00012c$ \cite{Bowlan09OL}, and $1.00015c$ \cite{Kuntz09PRA}, and $1.111c$ in a plasma \cite{Alexeev02PRL}. Reports on measured subluminal speeds have been lacking \cite{Turunen10PO} and limited to delays of hundreds of femtoseconds over a distance of 10~cm \cite{Lohmus12OL,Piksarv12OE}, corresponding to a group velocity of $\approx0.999c$. There have been no experimental reports to date on negative group velocities in free space.

Here, we synthesize `space-time' (ST) wave packets \cite{Kondakci16OE,Parker16OE,Kondakci17NP} using a phase-only spatial light modulator (SLM) that efficiently sculpts the field spatio-temporal spectrum and modifies the group velocity dynamically. The ST wave packets are synthesized for simplicity in the form of a light sheet that extends uniformly in one transverse dimension over $\sim25$~mm, such that control over $v_{\mathrm{g}}$ is exercised in a macroscopic volume of space. We measure $v_{\mathrm{g}}$ in an interferometric arrangement utilizing a reference pulsed plane wave and confirm precise control over $v_{\mathrm{g}}$ from $30c$ in the forward direction to $-4c$ in the backward direction. We observe group delays of $\sim\pm30$~ps (three orders-of-magnitude larger than those in \cite{Giovannini15Science,Bouchard16Optica}), which is an order-of-magnitude longer than the pulse width, and is observed over a distance of only $\sim10$~mm. Adding to the uniqueness of our approach, the achievable group velocity is \textit{independent} of the beam size and of the pulse width. All that is needed to change the group velocity is a reorganization of the spectral correlations underlying the wave packet spatio-temporal structure.

The novelty of our approach is its reliance on a linear system that utilizes a phase-only spatio-temporal Fourier synthesis strategy, which is energy-efficient and precisely controllable \cite{Kondakci17NP}. Our approach allows for endowing the field with arbitrary, programmable spatio-temporal spectral correlations that can be tuned to produce -- smoothly and continuously -- any desired wave packet group velocity. The precision of this technique with respect to previous approaches is attested by the unprecedented range of control over the measured group velocity values over the subluminal, superluminal, and negative regimes in a single optical configuration. Crucially, while distinct theoretical proposals have been made previously for each range of the group velocity (e.g., subluminal \cite{Liu98JMO,Sheppard02JOSAA,Zapata08JOSAA}, superluminal \cite{Valtna07OC}, and negative \cite{Zapata06OL} spans), our strategy is -- to the best of our knowledge -- the only experimental arrangement capable of controlling the group velocity continuously across all these regimes (with no moving parts) simply through the electronic implementation of a phase pattern imparted to a spectrally spread wave front impinging on a SLM.

\section*{Results}

\begin{figure}[b!]
\centering
\includegraphics[scale=1.0]{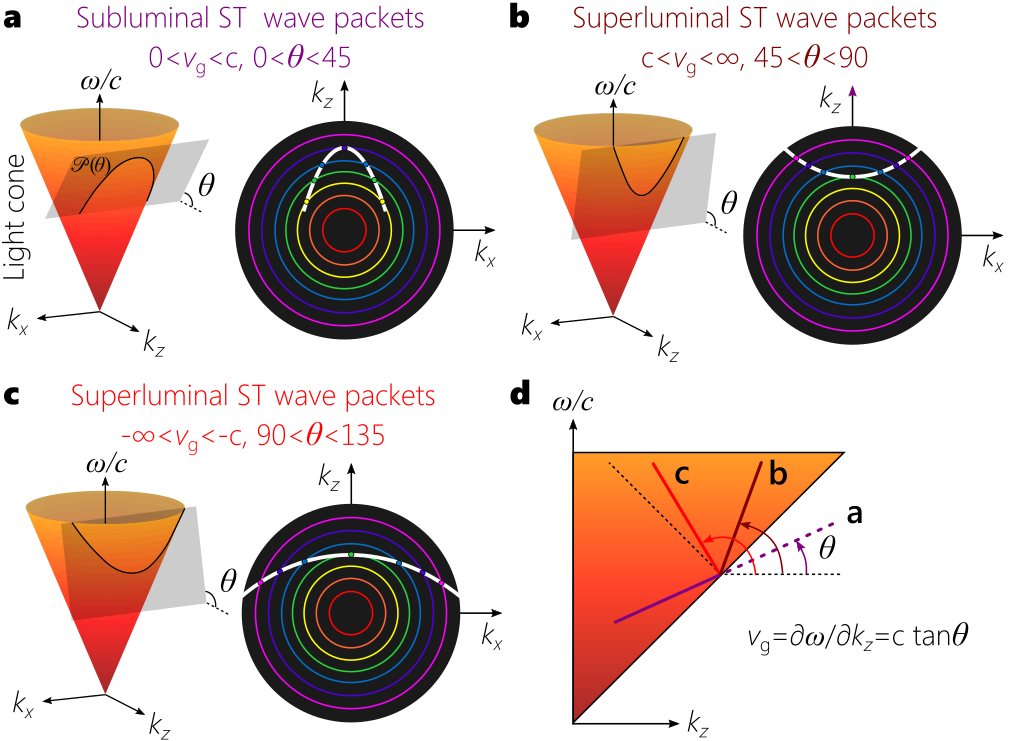} 
\caption{\textbf{Spatio-temporal spectral engineering for arbitrary control over the group velocity of an optical wave packet in free space.} \textbf{a-c}, Conic-section trajectories at the intersection of the light-cone with a spectral hyperplane $\mathcal{P}(\theta)$ are spatio-temporal spectral loci of ST wave packets with tunable group velocities $v_{\mathrm{g}}$ in free space: \textbf{a}, subluminal; \textbf{b}, superluminal; \textbf{c}, negative-superluminal $v_{\mathrm{g}}$. For each case we plot the projection of the spatio-temporal spectrum onto the $(k_{x},k_{z})$-plane restricted to $k_{z}>0$ (dashed white curve). \textbf{d}, Projections of the spatio-temporal spectra from (\textbf{a}-\textbf{c}) onto the $(k_{z},\tfrac{\omega}{c})$-plane. The slope of each projection determines the group velocity of each ST wave packet along the axial coordinate $z$, $v_\mathrm{g}\!=\!c\tan{\theta}$.}
\label{Fig:1}
\end{figure}

\subsection*{Concept of space-time wave packets.}

\vspace{-3mm}

\noindent 
The properties of ST light sheets can be best understood by examining their representation in terms of monochromatic plane waves $e^{i(k_{x}x+k_{z}z-\omega t)}$, which are subject to the dispersion relationship $k_{x}^{2}+k_{z}^{2}\!=\!(\tfrac{\omega}{c})^{2}$ in free space; here, $k_{x}$ and $k_{z}$ are the transverse and longitudinal components of the wave vector along the $x$ and $z$ coordinates, respectively, $\omega$ is the temporal frequency, and the field is uniform along $y$. This relationship corresponds geometrically in the spectral space $(k_{x},k_{z},\tfrac{\omega}{c})$ to the surface of the light-cone (Fig.~\ref{Fig:1}). The spatio-temporal spectrum of any physically realizable optical field compatible with causal excitation must lie on the surface of the light-cone with the added restriction $k_{z}\!>\!0$. For example, the spatial spectra of monochromatic beams lie along the circle at the intersection of the light-cone with a horizontal iso-frequency plane, whereas the spatio-temporal spectrum of a traditional pulsed beam occupies a two-dimensional (2D) patch on the light-cone surface.

The spectra of ST wave packets do \textit{not} occupy a 2D patch, but instead lie along a curved one-dimensional (1D) trajectory resulting from the intersection of the light-cone with a tilted spectral hyperplane $\mathcal{P}(\theta)$ described by the equation $\tfrac{\omega}{c}\!=\!k_{\mathrm{o}}+(k_{z}-k_{\mathrm{o}})\tan{\theta}$, where $k_{\mathrm{o}}\!=\!\tfrac{\omega_{\mathrm{o}}}{c}$ is a fixed wave number \cite{Kondakci17NP}, and the ST wave packet thus takes the form
\begin{eqnarray}\label{Eq:PlaneWaveRepresentation}
E(x,z;t)&=&e^{i(k_{\mathrm{o}}z-\omega_{\mathrm{o}}t)}\int\!dk_{x}\tilde{\psi}(k_{x})e^{i(k_{x}x+[k_{z}-k_{\mathrm{o}}][z-ct\tan{\theta}])}\nonumber\\
&=&e^{i(k_{\mathrm{o}}z-\omega_{\mathrm{o}}t)}\psi(x,z-v_{\mathrm{g}}t).
\end{eqnarray}
Therefore, the group velocity along the $z$-axis is 
$v_{\mathrm{g}}\!=\!\tfrac{\partial\omega}{\partial k_{z}}\!=\!c\tan{\theta}$, and is determined solely by the tilt of the hyperplane $\mathcal{P}(\theta)$. In the range $0\!<\!\theta\!<\!45^{\circ}$, we have a \textit{sub}luminal wave packet $v_{\mathrm{g}}\!<\!c$, and $\mathcal{P}(\theta)$ intersects with the light-cone in an ellipse [Fig.~\ref{Fig:1}a]. In the range $45^{\circ}\!<\!\theta\!<\!90^{\circ}$, we have a \textit{super}luminal wave packet $v_{\mathrm{g}}\!>\!c$, and $\mathcal{P}(\theta)$ intersects with the light-cone in a hyperbola [Fig.~\ref{Fig:1}b]. Further increasing $\theta$ reverses the \textit{sign} of $v_{\mathrm{g}}$ such that the wave packet travels \textit{backwards} towards the source $v_{\mathrm{g}}\!<\!0$ in the range $90^{\circ}\!<\!\theta\!<\!180^{\circ}$ [Fig.~\ref{Fig:1}c]. These various scenarios are summarized in Fig.~\ref{Fig:1}d.

\subsection*{Experimental realization}

\vspace{-3mm}

\begin{figure}[t!]
\centering
\includegraphics[scale=1.0]{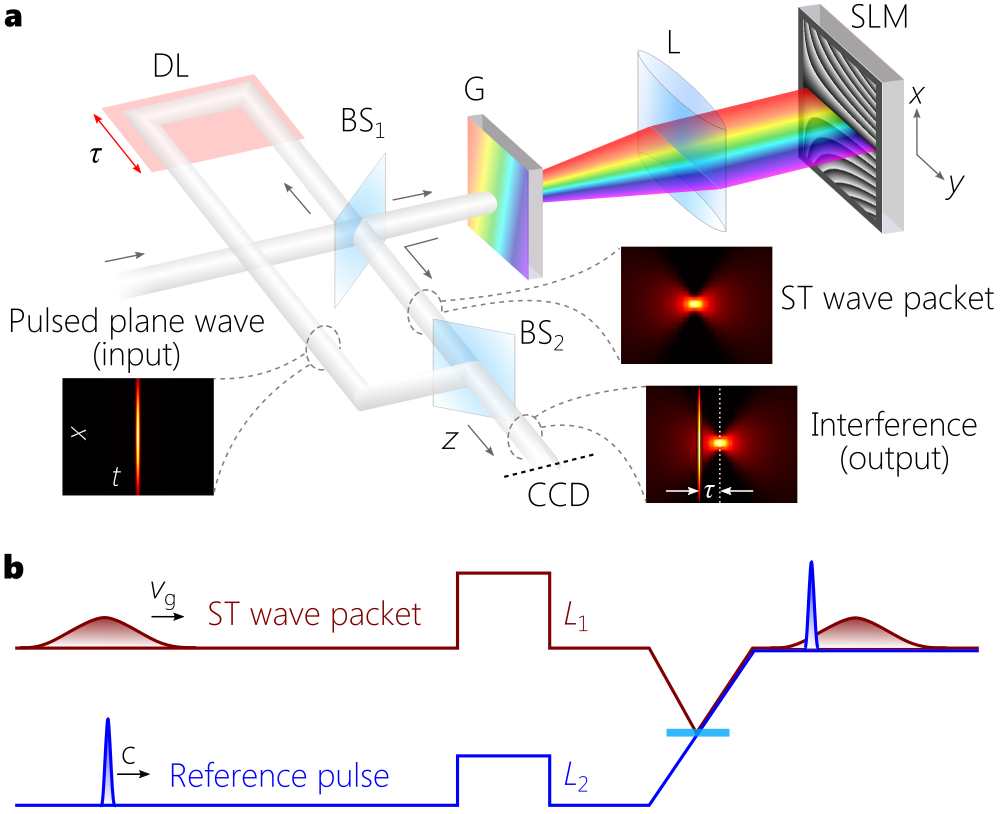} 
\caption{\textbf{Synthesizing ST wave packets and measuring their group velocity.} \textbf{a}, A pulsed plane wave is split into two paths: in one path the ST wave packet is synthesized using a two-dimensional pulse shaper formed of a diffraction grating (G) and spatial light modulator (SLM), while the other path is the reference. L: Cylindrical lens; BS: beam splitter; CCD: charge coupled device; DL: delay line. The insets provide the spatio-temporal profile of a ST wave packet with a Gaussian spectrum, the reference pulsed plane wave, and their interference. See Methods and Supplementary Material for details. \textbf{b}, The reference and the ST wave packets are superposed, and the shorter reference pulse probes a fraction of the longer ST wave packet. Maximal interference visibility is observed when the selected delays $L_{1}$ and $L_{2}$ cause their peaks to coincide; see Methods and Supplementary Material.}
\label{Fig:2}
\end{figure}

\begin{figure*}[ht!]
\centering
\includegraphics[scale=1]{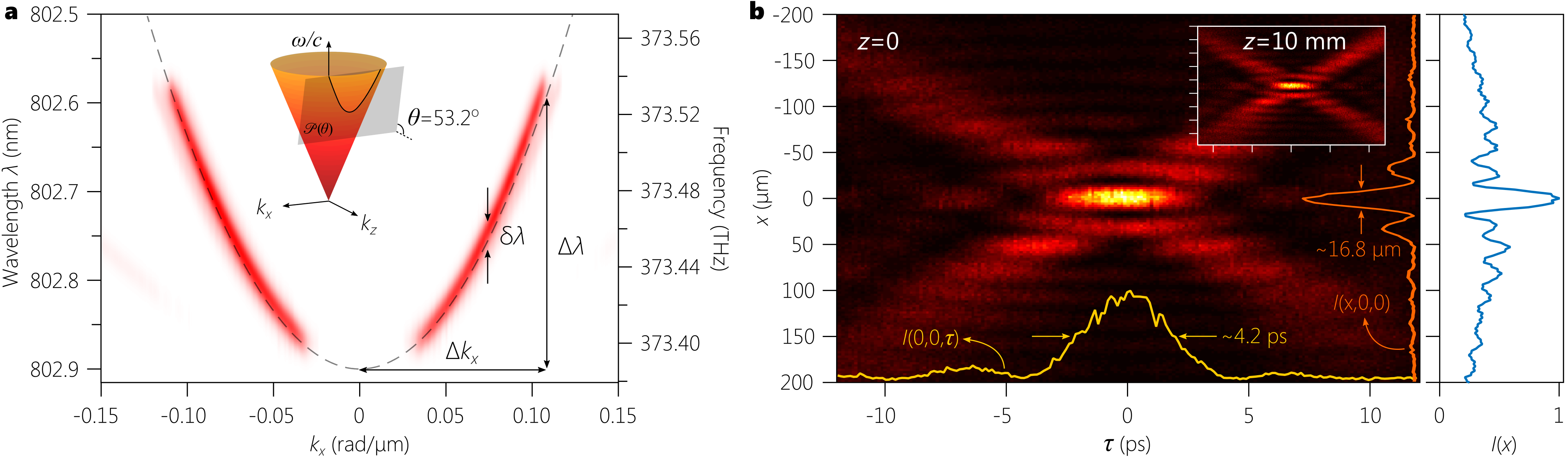}
\caption{\textbf{Spatio-temporal measurements of ST wave packets.} \textbf{a},\textbf{b}, Spatio-temporal (\textbf{a}) spectrum $|\tilde{E}(k_{x},\lambda)|^{2}$ and (\textbf{b}) intensity profile $I(x,0,\tau)$ for a ST wave packet with superluminal group velocity $v_\mathrm{g}=1.34c$, corresponding to a spectral hyperplane with $\theta=53.2^{\circ}$. \textbf{b}, The yellow and orange lines depict the pulse profile at $x=0$, $I(0,0,\tau)$, and beam profile at $\tau=0$, $I(x,0,0)$, respectively. The inset shows the spatio-temporal intensity profile after propagating for $z=10$~mm confirming the self-similar evolution of the ST wave packet. On the right, the normalized time-integrated beam profile $I(x,0)=\int d\tau I(x,0,\tau)$ is given.}
\label{Fig:3}
\end{figure*}

\noindent
We synthesize the ST wave packets by sculpting the spatio-temporal spectrum in the $(k_{x},\tfrac{\omega}{c})$-plane via a two-dimensional pulse shaper \cite{Kondakci17NP,Kondakci18PRL}. Starting with a generic pulsed plane wave, the spectrum is spread in space via a diffraction grating before impinging on a SLM, such that each wavelength $\lambda$ occupies a column of the SLM that imparts a linear phase corresponding to a pair of spatial frequencies $\pm k_{x}$ that are to be assigned to that particular wavelength, as illustrated in Fig.~\ref{Fig:2}a; see Methods. The retro-reflected wave front returns to the diffraction grating that superposes the wavelengths to reconstitute the pulse and produce the propagation-invariant ST wave packet corresponding to the desired hyperplane $\mathcal{P}(\theta)$. Using this approach we have synthesized and confirmed the spatio-temporal spectra of 11 different ST wave packets in the range $0^{\circ}<\theta<180^{\circ}$ extending from the subluminal to superluminal regimes. Figure~\ref{Fig:3} shows the measured spatio-temporal spectral intensity $|\tilde{E}(k_{x},\lambda)|^{2}$ (Fig.~\ref{Fig:3}a) for a ST wave packet having $\theta\!=\!53.2^{\circ}$ and thus lying on a hyperbolic curve on the light-cone corresponding to a positive superluminal group velocity of $v_{\mathrm{g}}\!=\!1.34c$. The spatial bandwidth is $\Delta k_{x}\!=\!0.11$~rad/$\mu$m, and the the temporal bandwidth is $\Delta\lambda\approx0.3$~nm. This spectrum is obtained by carrying out an optical Fourier transform along $x$ to reveal the spatial spectrum, and resolving the temporal spectrum with a diffraction grating. Our spatio-temporal synthesis strategy is distinct from previous approaches that make use of Bessel-beam-generation techniques and similar methodologies \cite{Saari97PRL,Alexeev02PRL,Valtna07OC,Bonaretti09OE,Bowlan09OL}, nonlinear processes \cite{DiTrapani03PRL,Faccio06PRL,Faccio07OE}, or spatio-temporal filtering \cite{Dallaire09OE,Jedrkiewicz13OE}. The latter approach utilizes a diffraction grating to spread the spectrum in space, a Fourier spatial filter then carves out the requisite spatio-temporal spectrum, resulting either in low throughput or high spectral uncertainty. In contrast, our strategy exploits a phase-only modulation scheme that is thus energy-efficient and can smoothly and continuously (within the precision of the SLM) tune the spatio-temporal correlations electronically with no moving parts, resulting in a corresponding controllable variation in the group velocity.

To map out the spatio-temporal profile of the ST wave packet $I(x,z,t)=|E(x,z,t)|^{2}$, we make use of the interferometric arrangement illustrated in Fig.~\ref{Fig:2}a. The initial pulsed plane wave (pulse width $\sim100$~fs) is used as a reference and travels along a delay line that contains a spatial filter to ensure a flat wave front (see Methods and Supplementary Material). Superposing the shorter reference pulse and the synthesized ST wave packet (Eq.~\ref{Eq:PlaneWaveRepresentation}) produces spatially resolved interference fringes when they overlap in space \textit{and} time -- whose visibility reveals the spatio-temporal pulse profile. The measured intensity profile $I(x,0,\tau)=|E(x,0,\tau)|^{2}$ of the ST wave packet having $\theta\!=\!53.2^{\circ}$ is plotted in Fig.~\ref{Fig:3}b; $\tau$ is the delay in the reference arm. Plotted also are the pulse profile at the beam center $I(0,0,\tau)=|E(x\!=\!0,0,\tau)|^{2}$ whose width is $\approx4.2$~ps, and the beam profile at the pulse center $I(x,0,0)=|E(x,0,\tau\!=\!0)|^{2}$ whose width is $\approx16.8$~$\mu$m. Previous approaches for mapping out the spatio-temporal profile of propagation-invariant wave packets have made use of strategies ranging from spatially resolved ultrafast pulse measurement techniques \cite{Bowlan09OL,Lohmus12OL,Piksarv12OE} to self-referenced interferometry \cite{Dallaire09OE,Kondakci17NP}.

\subsection*{Controlling the group velocity of a space-time wave packet}

\vspace{-3mm}

\noindent
We now proceed to make use of this interferometric arrangement to determine $v_{\mathrm{g}}$ of the ST wave packets as we vary the spectral tilt angle $\theta$. The setup enables synchronizing the ST wave packet with the luminal reference pulse while also uncovering any dispersion or reshaping in the ST wave packet with propagation. We first synchronize the ST wave packet with the reference pulse and take the central peak of the ST wave packet as the reference point in space and time for the subsequent measurement. An additional propagation distance $L_{1}$ is introduced into the path of the ST wave packet, corresponding to a group delay of $\tau_{\mathrm{ST}}\!=\!L_{1}/v_{\mathrm{g}}\gg\Delta\tau$ that is sufficient to eliminate any interference. We then determine the requisite distance $L_{2}$ to be inserted into the path of the reference pulse to produce a group delay $\tau_{\mathrm{r}}\!=\!L_{2}/c$ and regain the maximum interference visibility, which signifies that $\tau_{\mathrm{ST}}\!=\!\tau_{\mathrm{r}}$. The ratio of the distances $L_{1}$ and $L_{2}$ provides the ratio of the group velocity to the speed of light in vacuum $L_{1}/L_{2}\!=\!v_{\mathrm{g}}/c$.

In the subluminal case $v_{\mathrm{g}}\!<\!c$, we expect $L_{1}\!<\!L_{2}$; that is, the extra distance introduced into the path of the reference traveling at $c$ is larger than that placed in the path of the \textit{slower} ST wave packet. In the superluminal case $v_{\mathrm{g}}\!>\!c$, we have $L_{1}\!>\!L_{2}$ for similar reasons. When considering ST wave packets having negative-$v_{\mathrm{g}}$, inserting a delay $L_{1}$ in its path requires \textit{reducing} the initial length of the reference path by a distance $-L_{2}$ preceding the initial reference point, signifying that the ST wave packet is traveling \textit{backwards} towards the source. As an illustration, the inset in Fig.~\ref{Fig:3}b plots the same ST wave packet shown in the main panel of Fig.~\ref{Fig:3}b observed after propagating a distance of $L_{1}\!=\!10$~mm, which highlights the self-similarity of its free evolution \cite{Kondakci17NP}. The time axis is shifted by $\tau_{\mathrm{r}}\!\approx\!24.88$~ps, corresponding to $v_\mathrm{g}=(1.36\pm4\times10^{-4})c$, which is excellent agreement with the expected value of $v_{\mathrm{g}}=1.34c$. 

\begin{figure}[t!]
\centering
\includegraphics[scale=1.0]{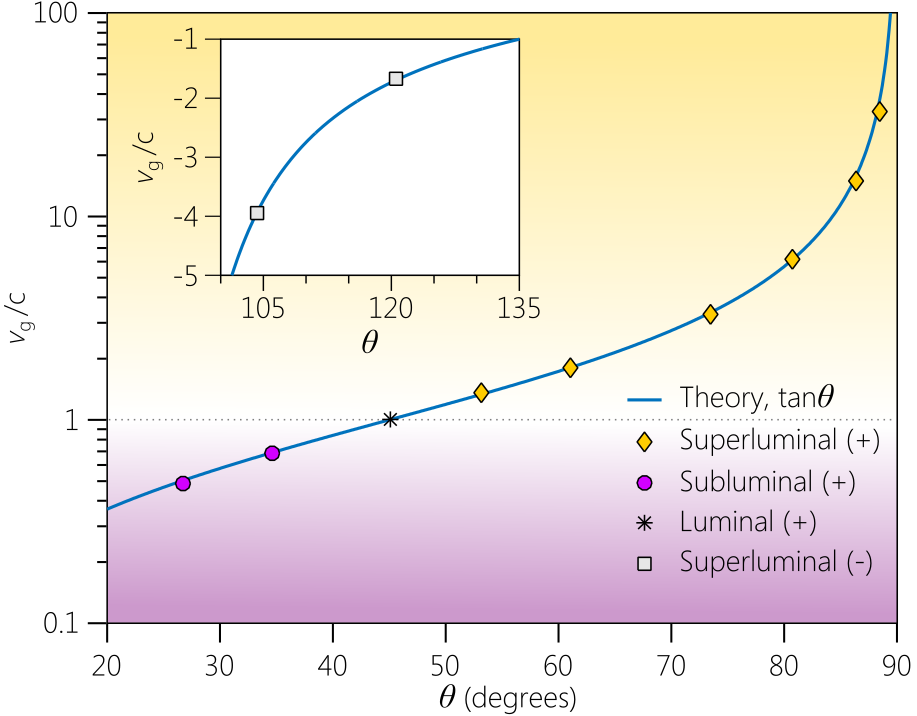} 
\caption{\textbf{Measured group velocities for ST wave packets.} By changing the tilt angle $\theta$ of the spectral hyperplane $\mathcal{P}$, we control $v_{\mathrm{g}}$ of the synthesized ST wave packets in the positive subluminal and superluminal regimes (corresponding to $0\!<\!\theta\!<\!90^{\circ}$). We plot $v_{\mathrm{g}}$ on a logarithmic scale. Measurements of ST wave packets with negative-$v_{\mathrm{g}}$ (corresponding to $\theta\!>\!90^{\circ}$) are given as points in the inset on a linear scale. The data in the main panel and in the inset is represented by points and both curves are the theoretical expectation $v_{\mathrm{g}}\!=\!c\tan{\theta}$. The error bars for the measurements are too small to appear, and are provided in Table~S1 in the Supplementary Material.}
\label{Fig:4}
\end{figure}

The results of measuring $v_{\mathrm{g}}$ while varying $\theta$ for the positive-$v_{\mathrm{g}}$ ST wave packets are plotted in Fig.~\ref{Fig:4}. The values of $v_{\mathrm{g}}$ range from subluminal values of $0.5c$ to the superluminal values extending up to $32c$ (corresponding to values of $\theta$ in the range $0^{\circ}\!<\!\theta\!<\!90^{\circ}$). The case of a \textit{luminal} ST wave packet corresponds trivially to a pulsed plane wave generated by idling the SLM. The data is in excellent agreement with the theoretical prediction of $v_{\mathrm{g}}\!=\!c\tan{\theta}$. The measurements of negative-$v_{\mathrm{g}}$ ($90^{\circ}\!<\!\theta\!<\!180^{\circ}$) are plotted in Fig.~\ref{Fig:4}, inset, down to $v_{\mathrm{g}}\!\approx\!-4c$, and once again are in excellent agreement with the expectation of $v_{\mathrm{g}}\!=\!c\tan{\theta}$. 

\section*{Discussion}

\noindent
An alternative understanding of these results makes use of the fact that tilting the plane $\mathcal{P}(\theta)$ from an initial position of $\mathcal{P}(0)$ corresponds to the action of a Lorentz boost associated with an observer moving at a relativistic speed of $v_{\mathrm{g}}=c\tan{\theta}$ with respect to a monochromatic source ($\theta=0$) \cite{Longhi04OE,Saari04PRE,Kondakci18PRL}. Such an observer perceives in lieu of the diverging monochromatic beam a non-diverging wave packet of group velocity $v_{\mathrm{g}}$ \cite{Belanger86JOSAA}. Indeed, at $\theta=90^{\circ}$ a condition known as `time-diffraction' is realized where the axial coordinate $z$ is replaced with time $t$, and the usual axial dynamics is displayed in time instead \cite{Longhi04OE,Porras17OL,Kondakci18PRL,PorrasPRA18}. In that regard, our reported results here on controlling $v_{\mathrm{g}}$ of ST wave packets is an example of relativistic optical transformations implemented in a laboratory through spatio-temporal spectral engineering.

Note that it is not possible in any finite configuration to achieve a delta-function correlation between each spatial frequency $k_{x}$ and wavelength $\lambda$; instead, there is always a finite spectral uncertainty $\delta\lambda$ in this association. In our experiment, $\delta\lambda\sim$24~pm (Fig.~\ref{Fig:3}a), which sets a limit on the diffraction-free propagation distance over which the modified group velocity can be observed \cite{Kondakci16OE}. The maximum group delay achieved is limited by the propagation-invariant length, which is dictated by the ratio of the temporal bandwidth $\Delta\lambda$ to the spectral uncertainty $\delta\lambda$. The finite system aperture ultimately sets the value of $\delta\lambda$. For example, the size of the diffraction grating determines its spectral resolving power, the finite pixel size of the SLM further sets a lower bound on the precision of association between the spatial and temporal frequencies, and the size of the SLM active area determines the maximum temporal bandwidth that can be exploited. Of course, the spectral tilt angle $\theta$ determines the proportionality between the spatial and temporal bandwidths, which then links these limits to the transverse beam width. The confluence of all these factors then determine the maximum propagation-invariant distance and hence the maximum achievable group delays. Careful design of the experimental parameters helps extend the propagation distance \cite{Bhaduri18OE}, and exploiting a phase plate in lieu of a SLM can extend the propagation distance even further \cite{Kondakci18OE}. Note that we have synthesized here optical wave packets where light has been localized along one transverse dimension but remains extended in the other transverse dimension. Localizing the wave packet along both transverse dimensions would require the addition of an additional SLM to extend the spatio-temporal modulation scheme into the second transverse dimension that we have not exploited here.

Finally, a different strategy has been recently proposed theoretically \cite{SaintMarie17Optica} and demonstrated experimentally \cite{Froula18NP} that makes use of a so-called `flying focus', whereupon a chirped pulse is focused with a lens having chromatic aberrations such that different spectral slices traverse the focal volume of the lens at a controllable speed, which was estimated by means of a streak camera. 

We have considered here ST wave packets whose spatio-temporal spectral projection onto the $(k_{z},\tfrac{\omega}{c})$-plane is a line. A plethora of alternative \textit{curved} projections may be readily implemented to explore different wave packet propagation dynamics and to accommodate the properties of material systems in which the ST wave packet travels. Our results pave the way to novel schemes for phase matching in nonlinear optical processes \cite{Averchi08PRA,Bahabad10NP}, new types of laser-plasma interactions \cite{Turnbull18PRL1,Turnbull18PRL2}, and photon-dressing of electronic quasiparticles \cite{Byrnes14NP}.


\section*{Methods}

\noindent
\textbf{Determining conic sections for the spatio-temporal spectra}

\vspace{2mm}

\noindent
The intersection of the light-cone $k_{x}^{2}+k_{z}^{2}\!=\!(\tfrac{\omega}{c})^{2}$ with the spectral hyperplane $\mathcal{P}(\theta)$ described by the equation $\tfrac{\omega}{c}\!=\!k_{\mathrm{o}}+(k_{z}-k_{\mathrm{o}})\tan{\theta}$ is a conic section: an ellipse ($0^{\circ}\!<\!\theta\!<\!45^{\circ}$ or $135^{\circ}\!<\!\theta\!<\!180^{\circ}$), a tangential line ($\theta\!=\!45^{\circ}$), a hyperbola ($45^{\circ}\!<\!\theta\!<\!135^{\circ}$), or a parabola ($\theta\!=\!135^{\circ}$). In all cases $v_{\mathrm{g}}\!=\!c\tan{\theta}$. The projection onto the $(k_{x},\tfrac{\omega}{c})$-plane, which the basis for our experimental synthesis procedure, is in all cases a conic section given by 
\begin{equation}
\frac{1}{k_{1}^{2}}\left(\tfrac{\omega}{c}\,\,\pm\,\,k_{2}\right)^{2}\pm\frac{k_{x}^{2}}{k_{3}^{2}}\!=\!1,
\end{equation}
where $k_{1}$, $k_{2}$ and $k_{3}$ are positive-valued constants: $\tfrac{k_{1}}{k_{\mathrm{o}}}\!=\!\left|\tfrac{\tan{\theta}}{1+\tan{\theta}}\right|$, $\tfrac{k_{2}}{k_{\mathrm{o}}}\!=\!\tfrac{1}{|1+\tan{\theta}|}$, and $\tfrac{k_{3}}{k_{\mathrm{o}}}\!=\!\sqrt{|\tfrac{1+\tan{\theta}}{1-\tan{\theta}}|}$. The signs in the equation are $(-,+)$ in the range $0\!<\!\theta\!<\!45^{\circ}$ (an ellipse), $(-,-)$ in the range $45^{\circ}\!<\!\theta\!<\!90^{\circ}$, and $(+,-)$ in the range $90^{\circ}\!<\!\theta\!<\!135^{\circ}$.

In the paraxial limit where $k_{x}^{\mathrm{max}}\!\ll\!k_{\mathrm{o}}$, the conic section in the vicinity of $k_{x}\!=\!0$ can be approximated by a section of a parabola, 
\begin{equation}
\frac{\omega}{\omega_{\mathrm{o}}}=1+f(\theta)\frac{k_{x}^{2}}{2k_{\mathrm{o}}^{2}},
\end{equation}
whose curvature is determined by $\theta$ through the function $f(\theta)$ given by
\begin{equation}
f(\theta)=\frac{\tan{\theta}}{\tan{\theta}-1}.
\end{equation}

\vspace{0.4cm}
\noindent
\textbf{Spatially resolved interferograms for resolving the spatio-temporal intensity profiles}

\vspace{2mm}

\noindent
We take the ST wave packet to be $E(x,z,t)\!=\!e^{i(k_{\mathrm{o}}z-\omega_{\mathrm{o}}t)}\psi(x,z-v_{\mathrm{g}}t)$ as provided in Eq.~\ref{Eq:PlaneWaveRepresentation}, and that of the reference plane-wave pulse to be $E_{\mathrm{r}}\!=\!e^{i(k_{\mathrm{o}}z-\omega_{\mathrm{o}}t)}\psi_{\mathrm{r}}(z-ct)$. We have dropped the $x$-dependence of the reference and $\psi_{\mathrm{r}}(z)$ is a slowly varying envelope. Superposing the two fields in the interferometer after delaying the reference by $\tau$ results in a new field $\propto E(x,z,t)+E_{\mathrm{r}}(x,z,t-\tau)$, whose time-average $I(x,\tau)$ is recorded at the output,  
\begin{equation}
I(x,\tau)\propto\int dt|E(x,z,t)+E_{\mathrm{r}}(x,z,t-\tau)|^{2}.
\end{equation}
We make use of the following representations of the fields for the ST wave packet and the reference pulse:
\begin{eqnarray}
E(x,z,t)&=&e^{i(k_{\mathrm{o}}z-\omega_{\mathrm{o}}t)}\int dk_{x}\tilde{\psi}(k_{x})e^{ik_{x}x}e^{-i(\omega-\omega_{\mathrm{o}})(t-{z}/{v_{\mathrm{g}}})}\nonumber\\
&=&e^{i(k_{\mathrm{o}}z-\omega_{\mathrm{o}}t)}\,\,\,\psi\left(x,t-{z}/{v_{\mathrm{g}}}\right),\\
E_{\mathrm{r}}(x,z,t)&=&e^{i(k_{\mathrm{o}}z-\omega_{\mathrm{o}}t)}\int d\omega\tilde{\psi}_{\mathrm{r}}(\omega-\omega_{\mathrm{o}})e^{i(\omega-\omega_{\mathrm{o}})(t-{z}/{c})}\nonumber\\
&=&e^{i(k_{\mathrm{o}}z-\omega_{\mathrm{o}}t)}\,\,\,\psi_{\mathrm{r}}\left(x,t-{z}/{c}\right).
\end{eqnarray}

We set the plane of the detector at $z\!=\!0$ (CCD$_1$ in our experiment; see Fig.~\ref{Fig:S1}), from which we obtain the spatio-temporal interferogram 
\begin{equation}
I(x,\tau)\propto I_{\mathrm{ST}}(x)+I_{\mathrm{r}}+2|R(x,\tau)|\cos{(\omega_{\mathrm{o}}\tau-\varphi_{\mathrm{R}}(x,\tau))},
\end{equation}
where
\begin{eqnarray}
I_{\mathrm{ST}}(x)&=&\int dt|\psi(x,t)|^{2}=\int dk_{x}|\tilde{\psi}(k_{x})|^{2}(1+\cos{2k_{x}x}),\\
I_{\mathrm{r}}&=&\int dt|\psi_{\mathrm{r}}(t)|^{2}=\int d\omega|\tilde{\psi}(\omega)|^{2},
\end{eqnarray}
where we have made the simplifying assumption that the spatial spectrum of the ST wave packet is an even function, $\tilde{\psi}(k_{x})\!=\!\tilde{\psi}(-k_{x})$. This assumption is applicable to our experiment and does not result in any loss of generality. Note that $I_{\mathrm{ST}}(x)$ corresponds to the time-averaged transverse spatial intensity profile of the ST wave packet, as would be registered by a CCD, for example, in absence of an interferometer. Similarly, $I_{\mathrm{r}}$ is equal to the time-averaged reference pulse and represents constant background term. Note that $z$ could be set at an arbitrary value because both the reference pulse and the ST wave packet are propagation-invariant.

The cross-correlation function $R(x,\tau)\!=\!|R(x,\tau)|e^{i\varphi_{\mathrm{R}}(x,\tau)}$ is given by
\begin{equation}
R(x,\tau)=\int dt\,\,\psi(x,t)\,\,\psi_{\mathrm{r}}^{*}(t-\tau).
\end{equation}
Taking the integral over time $t$ produces
\begin{equation}
R(x,\tau)=\int dk_{x}\tilde{\psi}(k_{x})\tilde{\psi}^{*}_{\mathrm{r}}(\omega)e^{ik_{x}x}e^{i(\omega-\omega_{\mathrm{r}})\tau},
\end{equation}
where $\omega$ is no longer an independent variable, but is correlate to the spatial frequency $k_{x}$ through the spatio-temporal curve at the intersection of the light-cone with the hyperspectral plane $\mathcal{P}(\theta)$. Because the reference pulse is significantly shorter that the ST wave packet, the spectral width of $\tilde{\psi}_{\mathrm{r}}$ is larger than that of $\tilde{\psi}$, so that one can ignore it, while retaining its amplitude,
\begin{eqnarray}
R(x,\tau)&\approx&|\tilde{\psi}_{\mathrm{r}}(\omega_{\mathrm{o}})|\int dk_{x}\tilde{\psi}(k_{x})e^{ik_{x}x}e^{i(\omega-\omega_{\mathrm{o}})\tau}\nonumber\\
&=&|\tilde{\psi}_{\mathrm{r}}(\omega_{\mathrm{o}})|\psi(x,\tau).
\end{eqnarray}
Note that the spectral function $\tilde{\psi}(k_{x})$ of the ST wave packet determines the coherence length of the observed spatio-temporal interferogram, which we thus expect to be on the order of the temporal width of the ST wave packet itself.

The visibility of the spatially resolved interference fringes is given by
\begin{equation}
\nu(x,\tau)\!=\!\frac{2|R(x,\tau)|}{I_{ST}(x)+I_{\mathrm{r}}}.
\end{equation}
The squared visibility is then given by
\begin{equation}
\nu^{2}(x,\tau)\!\approx\!\frac{4|\tilde{\psi}_{\mathrm{r}}(\omega_{\mathrm{o}})|^{2}|\psi(x,\tau)|^{2}}{I_{ST}(x)+I_{\mathrm{r}}}\propto|\psi(x,\tau)|^{2},
\end{equation}
where the last approximation requires that we can ignore $I_{\mathrm{ST}}(x)$ with respect to the constant background term $I_{\mathrm{r}}$ stemming from the reference pulse.

\vspace{0.4cm}
\noindent
\textbf{Details of the experimental setup}

\vspace{2mm}

\noindent
\textit{Synthesis of ST wave packets.}

\noindent
The input pulsed plane wave is produced by expanding the horizontally polarized pulses from a Ti:sapphire laser (Tsunami, Spectra Physics) having a bandwidth of $\sim8.5$~nm centered on a wavelength of 800~nm, corresponding to pulses having a width of $\sim100$~fs. A diffraction grating having a ruling of 1200~lines/mm and area $25\times25$~mm$^{2}$ in reflection mode (Newport 10HG1200-800-1) is used to spread the pulse spectrum in space and the second diffraction order is selected to increase the spectral resolving power, resulting in an estimated spectral uncertainty of $\delta\lambda\approx24$~pm. After spreading the full spectral bandwidth of the pulse in space, the width size of the SLM ($\approx16$~mm) acts as a spectral filter, thus reducing the bandwidth of the ST wave packet below the initial available bandwidth and minimizing the impact of any residual chirping in the input pulse. An aperture A can be used to further reduce the temporal bandwidth when needed. The spectrum is collimated using a cylindrical lens $L_{1-y}$ of focal length $f=50$~cm in a $2f$ configuration before impinging on the SLM. The SLM imparts a 2D phase modulation to the wave front that introduces controllable spatio-temporal spectral correlations. The retro-reflected wave from is then directed through the lens $L_{1-y}$ back to the grating G, whereupon the ST wave packet is formed once the temporal/spatial frequencies are superposed; see Fig.~\ref{Fig:S1}. Details of the synthesis procedure are described elsewhere \cite{Kondakci17NP,Kondakci18PRL,Kondakci18OE,Bhaduri18OE}.

\vspace{0.6cm}
\noindent
\textit{Spectral analysis of ST wave packets.}

\noindent
To obtain the spatio-temporal spectrum $|\tilde{E}(k_{x},\lambda)|^{2}$ plotted in Fig.~\ref{Fig:3}a in the main text, we place a beam splitter BS$_{2}$ within the ST synthesis system to sample a portion of the field retro-reflected from the SLM after passing through the lens L$_{1-y}$. The field is directed through a spherical lens L$_{4-\mathrm{s}}$ of focal length $f=7.5$~cm to a CCD camera (CCD$_{2}$); see Fig.~\ref{Fig:S1}. The distances are selected such that the field from the SLM undergoes a $4f$ configuration along the direction of the spread spectrum (such that the wavelengths remain separated at the plane of CCD$_2$), while undergoing a $2f$ system along the orthogonal direction, thus mapping each spatial frequency $k_{x}$ to a point.

\vspace{0.6cm}
\noindent
\textit{Reference pulse preparation.}

\noindent
The reference pulse is obtained from the initial pulsed beam before entering the ST wave packet synthesis stage via a beam splitter BS$_1$. The beam power is adjusted using a neutral density filter, and the spatial profile is enlarged by adding a spatial filtering system consisting of two lenses and a pinhole of diameter 30~$\mu$m. The spherical lenses are L$_{5-\mathrm{s}}$ of focal length $f=50$~cm and L$_{6-\mathrm{s}}$ of focal length $f=10$~cm, and they are arranged such that the pinhole lies at the Fourier plane. The spatially filtered pulsed reference then traverses an optical delay line before being brought together with the ST wave packet.

\vspace{0.6cm}
\noindent
\textit{Beam analysis.}

\noindent
The ST wave packet is imaged from the plane of the grating G to an output plane via a telescope system comprising two cylindrical lenses L$_{2-x}$ and L$_{3-x}$ of focal lengths 40~cm and 10~cm, respectively, arranged in a $4f$ system. This system introduced a demagnification by a factor $4\times$, which modifies the spatial spectrum of the ST wave packet. The phase pattern displayed by the SLM is adjusted to pre-compensate for this modification. The ST wave packet and the reference pulse are then combined into a common path via a beam splitter BS$_3$. A CCD camera (CCD$_{1}$) records the interference pattern resulting from the overlap of the ST wave packet and reference pulse, which takes place only when the two pulses overlap also in time; see Fig.~\ref{Fig:S3}. 

\vspace{0.4cm}
\noindent
\textbf{Details of group-velocity measurements}

\vspace{2mm}

\noindent
Moving CCD$_1$ a distance $\Delta z$ introduces an extra common distance in the path of both beams. However, since the ST wave packet travels at a group velocity $v_{\mathrm{g}}$ and the reference pulse at $c$, a relative group delay of $\Delta\tau\!=\!\Delta z(\tfrac{1}{c}-\tfrac{1}{v_{\mathrm{g}}})$ is introduced and the interference at CCD$_1$ is lost if $\Delta\tau\gg\Delta T$, where $\Delta T$ is the width of the ST wave packet in time. The delay line in the path of the reference pulse is then adjusted to introduce a delay $\tau\!=\!\Delta\tau$ to regain the interference. In the subluminal case $v_{\mathrm{g}}<c$, the reference pulse advances beyond the ST wave packet, and the interference is regained by increasing the delay traversed by the reference pulse with respect to the original position of the delay line. In the superluminal case $v_{\mathrm{g}}>c$, the ST wave packet advances beyond the reference pulse, and the interference is regained by reducing the delay traversed by the reference pulse with respect to the original position of the delay line. When $v_{\mathrm{g}}$ takes on negative values, the delay traversed by the reference pulse must be reduced even further. Of course, in the luminal case the visibility is not lost by introducing any extra common path distance $\Delta z$. See Fig.~\ref{Fig:S4} for a graphical depiction.

From this, the group velocity is given by
\begin{equation}
v_{\mathrm{g}}=\frac{\Delta z}{\Delta z/c-\Delta\tau}.
\end{equation}
For a given value of $\Delta z$, we fit the temporal profile $I(0,0,\tau)$ to a Gaussian function to determine its center from which we estimate $\Delta\tau$. For each tilt angle $\theta$, we repeat the measurement for three different values of $\Delta z$ and set one of the positions as the origin for the measurement set: 0~mm, 2~mm, and 4~mm in positive subluminal case (Fig.~\ref{Fig:S2}a); 0~mm, 5~mm, and 10~mm in the positive superluminal case (Fig.~\ref{Fig:S2}b); and 0~mm, -5~mm, and -10~mm in the negative-$v_{\mathrm{g}}$ case (Fig.~\ref{Fig:S2}c). Finally, we fit the obtained values to a linear function, where the slope corresponds to the group velocity. The uncertainty in estimating the values of $v_{\mathrm{g}}$ ($\Delta v_{\mathrm{g}}$ in Table~S1 and error bars for Fig.~\ref{Fig:4} in the main text) are obtained from the standard error in the slope resulting from the linear regression.



\def\bibsection{\section*{\refname}} 

\bibliography{diffraction}

\vspace{2mm}
\noindent
\textbf{Acknowledgments}\\
We thank D. N. Christodoulides and A. Keles for helpful discussions. This work was supported by the U.S. Office of Naval Research (ONR) under contract N00014-17-1-2458.

\newpage 

\renewcommand{\thepage}{S\arabic{page}}  
\renewcommand{\thesection}{S\arabic{section}}   
\renewcommand{\thetable}{S\arabic{table}}   
\renewcommand{\thefigure}{S\arabic{figure}}
\renewcommand{\theequation}{S\arabic{equation}}  

\setcounter{equation}{0}    
\setcounter{figure}{0}    
\setcounter{page}{1}

\onecolumngrid

\vspace*{15mm}
\section*{Supplementary Material}
\vspace*{5mm}

\begin{figure*}[htb]
\centering
\includegraphics[scale=1.3]{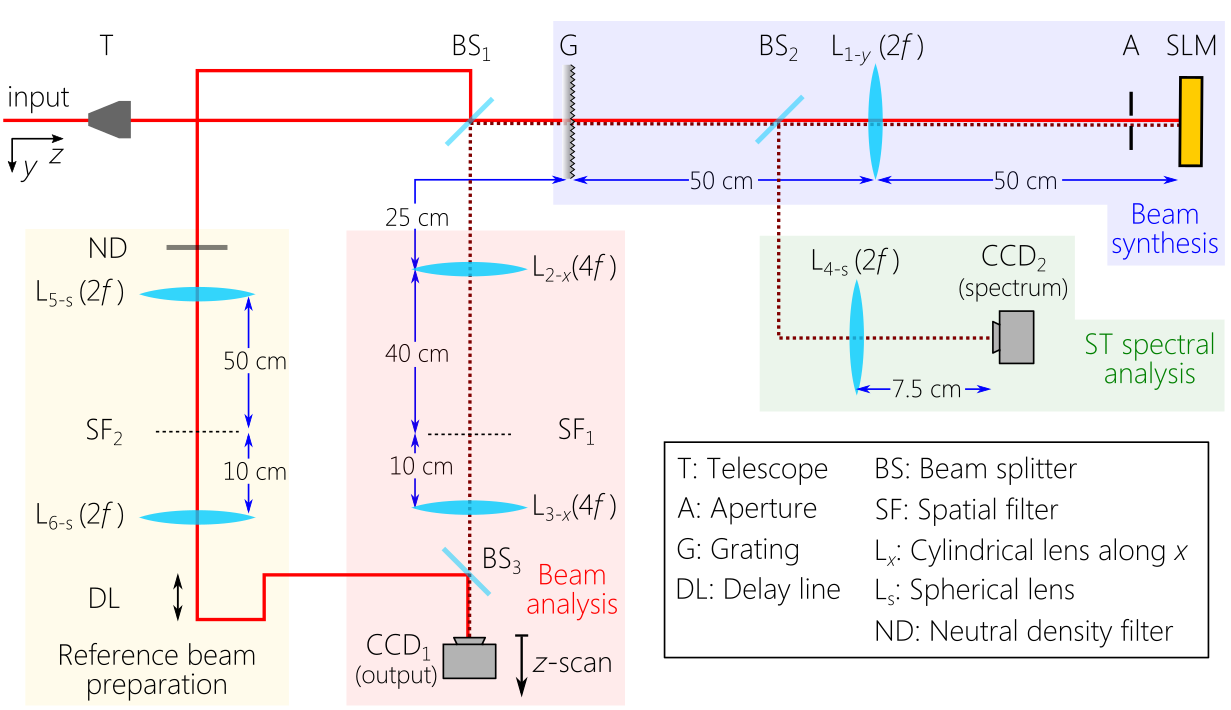} 
\caption{\textbf{Detailed experimental setup.} The experimental arrangement shown schematically in Fig.~\ref{Fig:2}a in the main text is presented here in detail. The setup comprises four sections for ST wave packet synthesis and characterization. The acronyms on all the optical components are provided in the inset box, and the focal lengths of the lenses and details of the experiment are provided in the text.}
\label{Fig:S1}
\end{figure*}

\begin{figure*}[htb]
\centering
\includegraphics[scale=0.8]{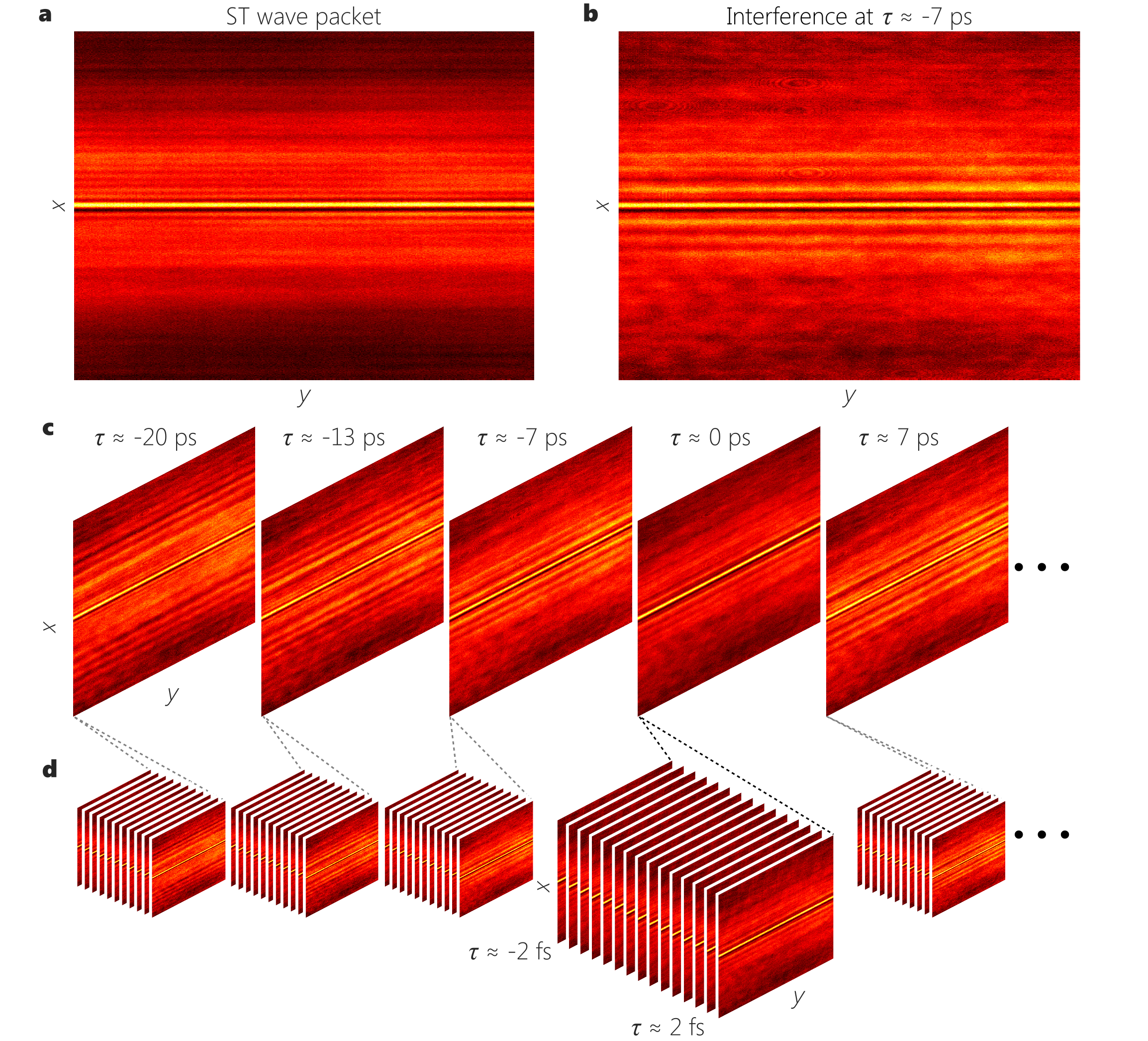} 
\caption{\textbf{Detailed procedure for restructuring the spatio-temporal wave packet profile from interferometric measurements.}  \textbf{a}, Time-integrated beam profile of a ST wave packet recorded by a slow detector (CCD$_1$) at a fixed propagation distance. In absence of the reference or if the reference and ST wave packet do not overlap in time, off-axis spatial interference along $x$ is absent. \textbf{b}, Spatially-resolved interference fringes along the $x$-axis for a delay $\tau\approx-7$~ps, resulting from the overlap of the reference and ST wave packet in time. \textbf{c}, The visibility of the spatially resolved interference fringes changes as the delay is scanned around the center of the ST wave packet. \textbf{d}, A set of measurements with small delay increments are taken to obtain the visibility $\nu$ near the vicinity of any selected $\tau$. Each set of such measurements yields a single line in the spatio-temporal profile $I(x,0,\tau)$ given in Fig.~\ref{Fig:2}b in the main text.}
\label{Fig:S3}
\end{figure*}

\begin{figure*}[htb]
\centering
\includegraphics[scale=1.3]{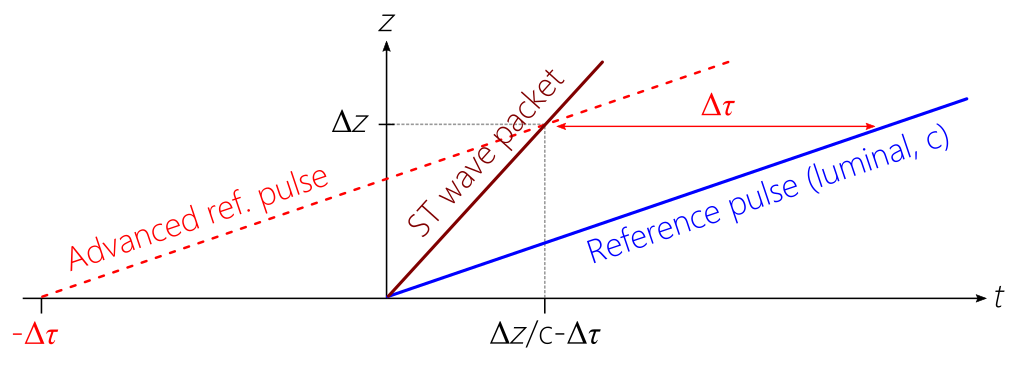} 
\caption{\textbf{Space-time-diagram for determining the group velocity of ST wave packets.} The experiment starts with the ST wave packet and reference pulse overlapping in space and time, resulting in high-visibility spatially resolved fringes (Fig.~\ref{Fig:S3}b). A common distance $\Delta z$ is introduced into the path of the ST wave packet and the reference pulse by moving CCD$_1$ (Fig.~\ref{Fig:S1}), which results in a loss of the interference fringes (Fig.~\ref{Fig:S3}a). A delay $\Delta\tau$ is then introduced into the path of the reference pulse to regain the visibility of the spatial resolved interference fringes (Fig.~\ref{Fig:S3}c).} 
\label{Fig:S4}
\end{figure*}

\begin{figure*}[htb]
\centering
\includegraphics[scale=0.8]{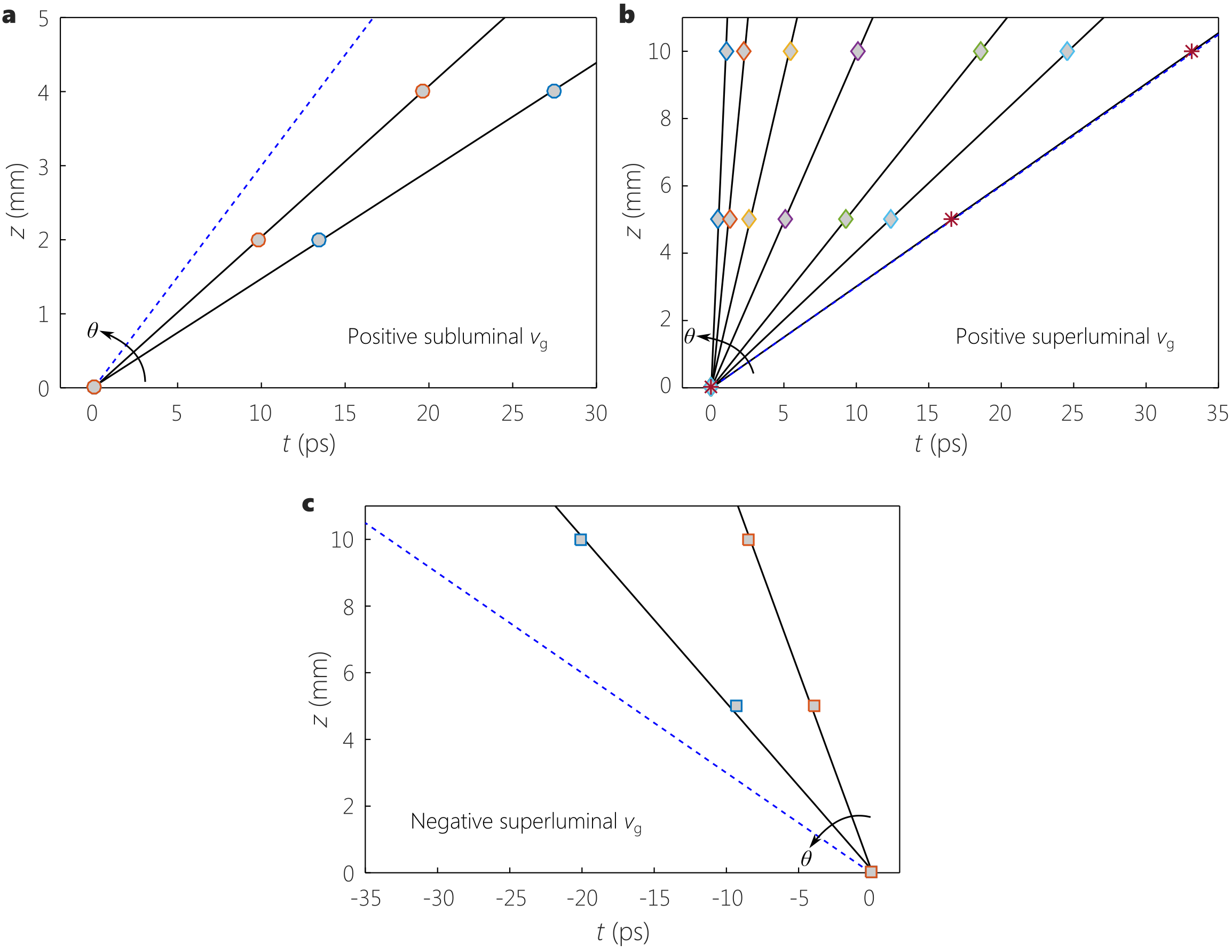} 
\caption{\textbf{Space-time-diagram measurements of ST wave packets to estimate their group velocities.} \textbf{a}, Positive subliminal ST wave packets lie below the light line $v_{\mathrm{g}}=c$ (blue-dashed). A retardation of the reference pulse is required to obtain maximum visibility. 
\textbf{b}, Positive superluminal ST wave packets lie above the light line with a positive slope. An advancement of the reference pulse is required to obtain maximum visibility. The luminal ST wave packet, which coincides -- as expected -- with the light line (data points represented by stars),  is produced by idling the SLM.  
\textbf{c}, Negative superluminal ST wave packets lie above the light line with a negative slope $v_{\mathrm{g}}=-c$. An advancement of the reference pulse is required to obtain maximum visibility. Note that the required advancement here exceeds that of the case of positive superluminal ST wave packets. 
\textbf{a-c}, In all cases, three measurements are taken at three points in $z$ and the required advancement or retardation of the reference pulse that gives the maximum visibility is recorded. One of the observation points set to be origin ($z=0$). The black lines are linear fits.} 
\label{Fig:S2}
\end{figure*}

{\renewcommand{\arraystretch}{0.8}
\begin{table*}[htb]\label{Table:GroupVelocityMeasurements}
  \centering
  \caption{Measurement results and theoretical expectation for the group velocity $v_{\mathrm{g}}$ of ST wave packets in free space, arranged in order of increasing value of $\theta$ in the range $0\!<\!\theta\!<\!180^{\circ}$. The theoretical values correspond to $v_{\mathrm{g}}=\tan{\theta}$, and $\Delta v_{\mathrm{g}}$ is the uncertainty in the measured value of $v_{\mathrm{g}}$.}
  \vspace{0.4cm}
  \begin{tabular}{p{0.3in}p{0.5in}p{1.7in}p{0.8in}p{1.0in}p{0.8in}p{1.0in}} 
  \hline 
 &\vspace{1mm} \textbf{$\theta$} & \vspace{1mm}\textbf{Wave packet type} & \vspace{1mm}\textbf{$v_{\mathrm{g}}$} & \vspace{1mm}\textbf{$\Delta v_{\mathrm{g}}$} & \vspace{1mm}\textbf{Theory} & \vspace{1mm}\textbf{Conic section} \\ \\ \hline

\vspace{0.05cm}&&&&&\\
(1) & $26.6^{\circ}$ & Positive subluminal & $0.49c$ & $\pm 3\times 10^{-4}c$ & $0.50c$ & ellipse\\
&&&&&\\\hline

&&&&&\\
(2) & $34.6^{\circ}$ & Positive subluminal & $0.68c$ & $\pm 2\times 10^{-5}c$ & $0.69c$ & ellipse\\
&&&&&\\\hline

&&&&&\\
(3) & $45^{\circ}$ & Positive luminal & $c$ & $\pm 3\times 10^{-5}c$ & $c$ & line\\
&&&&&\\\hline

&&&&&\\
(4) & $53.3^{\circ}$ & Positive superluminal & $1.36c$ & $\pm 4\times 10^{-4}c$ & $1.34c$ & hyperbola\\
&&&&&\\\hline

&&&&&\\
(5) & $61.1^{\circ}$ & Positive superluminal & $1.80c$ & $\pm 3\times 10^{-4}c$ & $1.81c$ & hyperbola\\
&&&&&\\\hline

&&&&&\\
(6) & $73.4^{\circ}$ & Positive superluminal & $3.29c$ & $\pm 2\times 10^{-4}c$ & $3.36c$ & hyperbola\\
&&&&&\\\hline

&&&&&\\
(7) & $80.7^{\circ}$ & Positive superluminal & $6.17c$ & $\pm 0.02c$ & $6.14c$ & hyperbola\\
&&&&&\\\hline

&&&&&\\
(8) & $86.4^{\circ}$ & Positive superluminal & $14.86c$ & $\pm 1.07c$ & $15.9c$ & hyperbola\\
&&&&&\\\hline

&&&&&\\
(9) & $88.5^{\circ}$ & Positive superluminal & $32.86c$ & $\pm 2.18c$ & $39.21c$ & hyperbola\\
&&&&&\\\hline

&&&&&\\
(10) & $104.2^{\circ}$ & Negative superluminal & $-3.94c$ & $\pm 0.03c$ & $-3.94c$ & hyperbola\\
&&&&&\\\hline

&&&&&\\
(11) & $120.6^{\circ}$ & Negative superluminal & $-1.66c$ & $\pm 5\times 10^{-3}c$ & $-1.69c$ & hyperbola\\
&&&&&\\\hline

\end{tabular}\label{TableSummary}
\end{table*}

\end{document}